\documentclass[aps,pre,twocolumn,showpacs]{revtex4}
\usepackage{epsfig}
\usepackage{graphicx}
\usepackage{bm}% bold math
\usepackage{amssymb}
\usepackage{amsmath}
\usepackage{amsthm}
\begin{document}
\title{Avalanches and clusters in planar crack front propagation}
\author{Lasse Laurson$^1$, Stephane Santucci$^2$, and Stefano Zapperi$^{1,3}$} 
\affiliation{$^1$ISI Foundation, Viale S. Severo 65, 10133 Torino, Italy}
\affiliation{$^2$Laboratoire de physique, CNRS UMR 5672, Ecole Normale Sup\'erieure de
Lyon, 46 All\'ee d'Italie, 69364 Lyon Cedex 07, France}
\affiliation{$^3$IENI-CNR, Via R. Cozzi 53, 20125, Milano, Italy}
\begin{abstract}
We study avalanches in a model for a planar crack propagating in a disordered
medium. Due to long-range interactions, avalanches are formed by a set of 
spatially disconnected local clusters, the sizes of which are distributed according 
to a power law with an exponent $\tau_{a}=1.5$. 
We derive a scaling relation $\tau_a=2\tau-1$ between the local cluster exponent 
$\tau_a$ and the global avalanche exponent $\tau$. For length scales longer than a
cross-over length proportional to the Larkin length, the aspect ratio of the local
clusters scales with the roughness exponent of the line model. Our analysis provides an 
explanation for experimental results on planar crack avalanches in Plexiglas plates, 
but the results are applicable also to other systems with long-range interactions. 
\end{abstract}
\pacs{62.20.mt, 45.70.Ht, 64.60.av}
\maketitle

\section{Introduction}

Driven elastic manifolds in disordered media have been used to model a
number of physical systems ranging from crack propagation in solids 
\cite{GAO-89,SCH-97,DEL-99,SCH-95,RAM-97,MAL-06,BON-08,SAL-06} to 
dynamics of magnetic domain 
walls in ferromagnets \cite{DUR-06} and vortices in type-II 
superconductors \cite{BLA-94}, and to charge density waves \cite{GRU-88}. 
Such systems exhibit non-equilibrium phase transitions as an 
external driving force $f$ is varied, so that below the depinning 
threshold $f_c$ the system is pinned by the disorder, while for $f>f_c$ 
the system moves at a finite average velocity. For $f$ close to but above 
$f_c$, such motion typically occurs as a sequence of avalanches with a 
broad distribution of sizes. A large class of such systems, including 
advancing crack fronts in solids \cite{GAO-89,SCH-97,DEL-99,MAL-06,BON-08,
SAL-06}, contact lines in wetting \cite{JOA-84}, magnetic 
domain walls with dipolar interactions \cite{DUR-06}, and 
plastically deforming crystals \cite{ZAI-06}, 
is characterized by a long-range interaction kernel. The peculiarity of these 
systems is that due to the long-range interactions governing the avalanche
dynamics, the area swept over by an avalanche is not necessarily 
simply connected in space as it would be the case if the interactions 
were only short-ranged. Instead, {\it global avalanches} are formed by 
a set of spatially disconnected {\it local clusters}.

The propagation of a planar crack confined in a weak plane represents
an ideal system to realize experimentally the depinning transition 
predicted by the crack line model \cite{GAO-89,SCH-95,RAM-97}. Yet,
the first experimental results for the crack front roughness \cite{SCH-97,DEL-99} 
and for the avalanche size distribution \cite{MAL-06} disagreed with the theory.
The roughness exponent of the crack front was found in the range 
$\zeta=0.55-0.6$ \cite{SCH-97,DEL-99}, instead of the theoretical value 
$\zeta =0.39$ \cite{ROS-02}, while the avalanche sizes defined by the waiting 
time matrix \cite{MAL-06}, measuring only spatially connected parts 
of the avalanches, were found experimentally to be power law 
distributed with an exponent much higher than the value predicted 
by the long-range line model for avalanches extracted by considering the global 
velocity signal of the crack front. These earlier claims have been revised 
recently: by testing a larger range of length scales it was shown that
the roughness exponent would cross over to the theoretical value at 
sufficiently large scales \cite{SAN-09,SAN-10}. Furthermore, in Ref.~\cite{BON-08} 
it was shown that the crack line model can reproduce the avalanche 
statistics of the experiment, if also the numerical data are analyzed 
using the waiting time matrix introduced in \cite{MAL-06}, while 
a smaller exponent value is obtained when considering avalanches 
extracted from the global velocity signal of the crack front model. 
Still, a complete understanding of the origin of the different scaling 
exponents is lacking.

Here we clarify these issues by studying the statistics of global avalanches 
and their spatially connected parts, the local avalanches (or clusters), 
for a long range elastic string moving in a disordered medium under 
{\it quasi-static} external driving. Both the avalanche and cluster size 
distributions 
are found to exhibit scaling, but with different power law exponents. In particular,
by studying the model with a larger range of length scales, we obtain a more 
accurate description of the scaling behavior, and therefore a better estimate
of the related exponents than in previous numerical studies \cite{BON-08}. 
These numerical results are compared with data from experiments on 
Plexiglas plates \cite{MAL-06}, and excellent agreement is found. Moreover,
we propose a scaling relation between the power law exponents
of the avalanche and cluster size distributions by considering the avalanche 
break-up process, supported by our numerical results. We finally study
the morphology of the local clusters, and find that their aspect ratio 
scales with the roughness exponent of the line model above a cross-over 
length proportional to the Larkin length
of the crack front. Previous studies \cite{BON-08} using a continuous 
time version of the crack line model have presumably been sampling the 
short length scale regime. Thus, our results explain and clarify a number of open 
issues related in particular to planar crack front propagation, such as 
the different scaling of global avalanches and local clusters, as well as
the connection between roughness and avalanche morphology. However, the implications
of our study extend beyond crack front propagation, as the results can be
immediately extended to a number of other systems described by the same model,
such as contact line dynamics in wetting \cite{JOA-84}, and low angle grain boundaries in
crystalline solids \cite{MOR-04}. Furthermore, a number of other driven systems in which the 
avalanche dynamics is governed by long range interactions, ranging from domain walls
in ferromagnets with dipolar interactions \cite{DUR-06} to plastically deforming crystalline
solids \cite{ZAI-06}, are expected to obey similar scaling laws. The paper is organized as follows:
In the next Section, the simulation model and the methods used to extract the local
clusters from the simulations and the experiments are presented. These are then 
followed by the results of the numerical simulations along with a direct comparison 
with experimental results from Plexiglas plates in Section III. Section IV finishes 
the paper with conclusions and discussion.  

\section{Model and methods}

\subsection{Quasi-static model of crack front propagation}

The model of the propagating planar crack front, introduced in Ref.~\cite{SCH-95}, 
is represented by a vector of integer heights $h_i$, $i=1,\dots,L$, with $L$ the 
system size. Crack propagation is driven by the local stress intensity factor (SIF) 
$K_i$, which represents the asymptotic prefactor of the $1/\sqrt{r}$ divergence of 
the stress field near the crack tip. To model 
the different contributions of the local SIF $K_i$ acting on a front element $i$, it is taken
to be of the form  $K_i=K_i^{elastic}+K_{i,h_i}^{random}+K^{ext}$, where 
\begin{equation}
\label{eq:kernel}
K_i^{elastic} = \Gamma_0 \sum_{j \neq i}^L \frac{h_j-h_i}{b|j-i|^2}
\end{equation}
represents the first order variation of the stress intensity factor
due to a first-order perturbation of the front position, $b$ is the front segment spacing 
and $\Gamma_0$ tunes the strength of the elastic interactions \cite{GAO-89}, 
$K_{i,h_i}^{random}$ is a time-independent Gaussian random variable of zero mean and 
unit variance, with 
\begin{equation}
\langle K_{i,h_i}^{random} K_{j,h_j}^{random} \rangle = \delta(i-j)\delta(h_i-h_j),
\end{equation} 
representing random  toughness of the material, 
and $K^{ext}$ is the contribution of the external load. Notice that the long range 
kernel in Eq.~(\ref{eq:kernel}) is identical in many other problems, such
as contact lines \cite{JOA-84} and low angle grain boundaries \cite{MOR-04}. 
Periodic boundary conditions are imposed. The dynamics is defined in discrete 
time $t$ by setting
\begin{equation}
v_i(t)=h_i(t+1)-h_i(t)=\theta(K_i),
\end{equation}
where $v_i$ is the local velocity of the front element $i$, and 
$\theta$ is the Heaviside step function. During a single time
step, the front elements with $v_i>0$ are advanced by a unit step,
$h_i(t+1)= h_i(t)+1$, and new random forces are generated for those 
elements. The local forces are then computed again for each 
element, and the process is repeated until $v_i=0$ for all $i$
and the avalanche stops. The size $s$ of the avalanche is the total
number of elementary moves during such an avalanche. 
Then the external stress 
is increased so that exactly one of the elements becomes unstable, and a 
new avalanche is initiated. 
While such a discretization of the dynamics neglects the fact that the
local velocity of the crack front should be proportional to the local SIF,
this is a standard technique employed in models of the depinning transition,
going back to the work of Leschhorn \cite{LES-93}, and is known to
have no influence on the scaling behavior we study here. Moreover,
this procedure is essential to be able to drive the system quasi-statically.
Such quasi-static driving has the advantage over continuous time models
with a finite driving velocity \cite{BON-08} that avalanches and clusters
can be defined without ambiguity also for a zero threshold.
%Notice that such quasistatic driving implies that avalanches can 
%be defined without the need to threshold the activity. 
As the crack front advances, the applied SIF $K^{ext}$ 
decreases at a rate proportional to the instantaneous average velocity 
$v(t)=1/L\sum_{i=1}^L v_i(t)$ of the front, with a proportionality 
constant $k$. Thus, the cut-off of the avalanche size distributions is 
expected to scale with $k$ \cite{DUR-06}.

\begin{figure}[t]
\begin{center}
\includegraphics[width=8.0cm]
{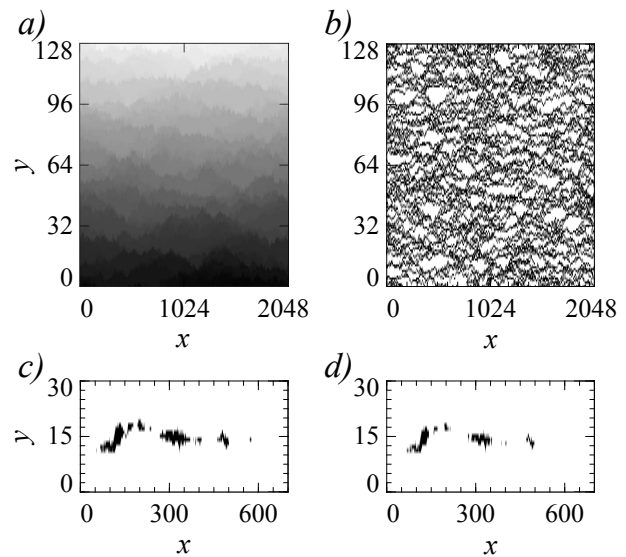}
\caption{{\bf a)} An example of the avalanche map $A(x,y)$, for $k=0.05$
and $\Gamma_0=1$, with the area swept over by different avalanches denoted 
by different shades of gray (time is running from black to white). 
{\bf b)} The same data as in {\bf a)} represented by the velocity
matrix $V(x,y)=1/W(x,y)$ with a zero threshold. The crack front is
moving to the positive $y$ direction. Black corresponds to zero velocity,
while regions of finite velocity are shown in white. {\bf c)} An
example of a structure of a single avalanche as identified by the AM
method. {\bf d)} The same avalanche as in {\bf c)} extracted
by using the WTM method.}
\label{fig:maps}
\end{center}
\end{figure}

\subsection{Methods to extract the local clusters}

To extract the clusters from an avalanche, we use two methods. The first is
the avalanche map (AM), based on a
two dimensional array $A(x,y)$ such that a unique value is assigned to all 
points $(x,y)$ over which an 
avalanche has swept. We then define the local cluster size 
$a^{AM}$ as the area of spatially connected regions with the
same value of $A(x,y)$ (see Fig. 
\ref{fig:maps} a and c). The second approach is the
recently proposed waiting time matrix (WTM) $W(x,y)$, 
defined by the time the interface has spent within a pixel corresponding 
to the location $(x,y)$ \cite{MAL-06,BON-08}. The local velocity matrix
is then given by $V(x,y)=1/W(x,y)$. Clusters are
defined as spatially connected regions of area $a^{WTM}$ within which the local velocity
exceeds some threshold value $v_{th}$ (see Fig. \ref{fig:maps} b and d). 
Here we study the model in the quasi-static limit, which allows us to use 
effectively a zero threshold, by setting the waiting time of the pinned 
configurations separating avalanches to be much 
longer than the maximum avalanche duration.
When using the WTM approach to record the motion of the crack front during an avalanche,
only $n-1$ steps are recorded when a line element moves $n$ steps, as 
both the initial and final pinned configurations are classified as immobile. 
Thus, the avalanche in Fig. \ref{fig:maps} d) is a bit thinner in the $y$ 
direction as compared to the one in Fig. \ref{fig:maps} c), and parts of 
the avalanche involving motion of a single step forward only are lost 
completely. However, our results suggest that this does not affect the statistical 
properties of the clusters. In the case of the discrete line model, 
another issue is that for soft lines ($\Gamma_0<0.75$), in which the 
height difference between neighboring segments of the line can be two or 
more pixels of the WTM, two consecutive avalanches can be recorded as one. 
In the following we use WTM only for large enough $\Gamma_0$. 
Notice that the AM does not suffer from such problems.

\begin{figure}[t]
\begin{center}
\includegraphics[width=6.5cm,angle=-90,clip]           
{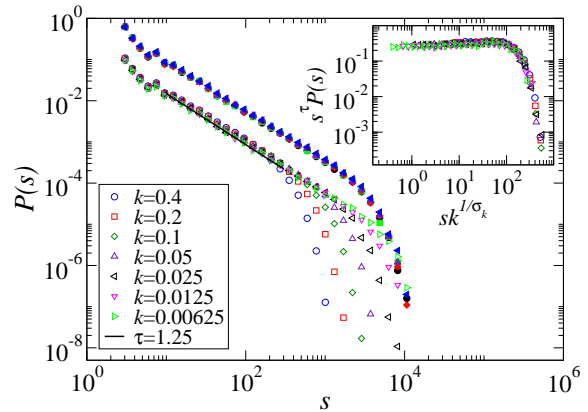}
\caption{(color online)
The main figure shows the size distributions of the avalanches
for $\Gamma_0=1$ and various values of $k$ (open symbols),
as well as the size distributions of the avalanches for a 
fixed $k=0.0125$ and various values of $\Gamma_0$ ranging
from $\Gamma_0=0.4$ to $\Gamma_0=1.0$ (filled symbols, with the
latter set of distributions displaced vertically for clarity). 
The solid line is a guide to the eye and corresponds to 
$\tau=1.25$. The inset shows a data collapse of the distributions
with $\Gamma_0=1$ and various values of $k$, with $\tau=1.25$ 
and $1/\sigma_k=0.725$.}
\label{fig:global_dists}
\end{center}
\end{figure}

\begin{figure}[t]
\begin{center}
\includegraphics[width=6.5cm,angle=-90,clip]
{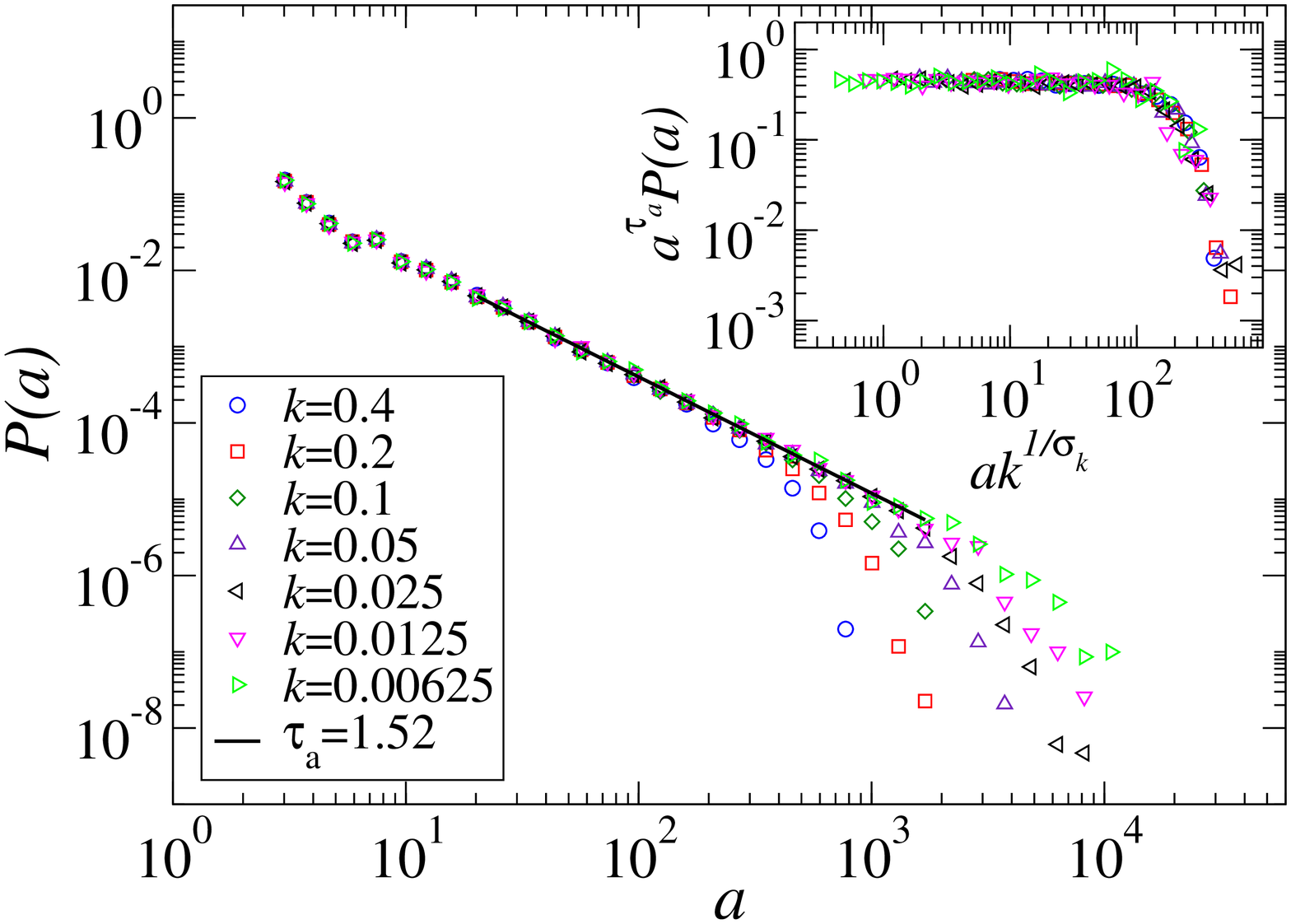}
\includegraphics[width=6.5cm,angle=-90,clip]
{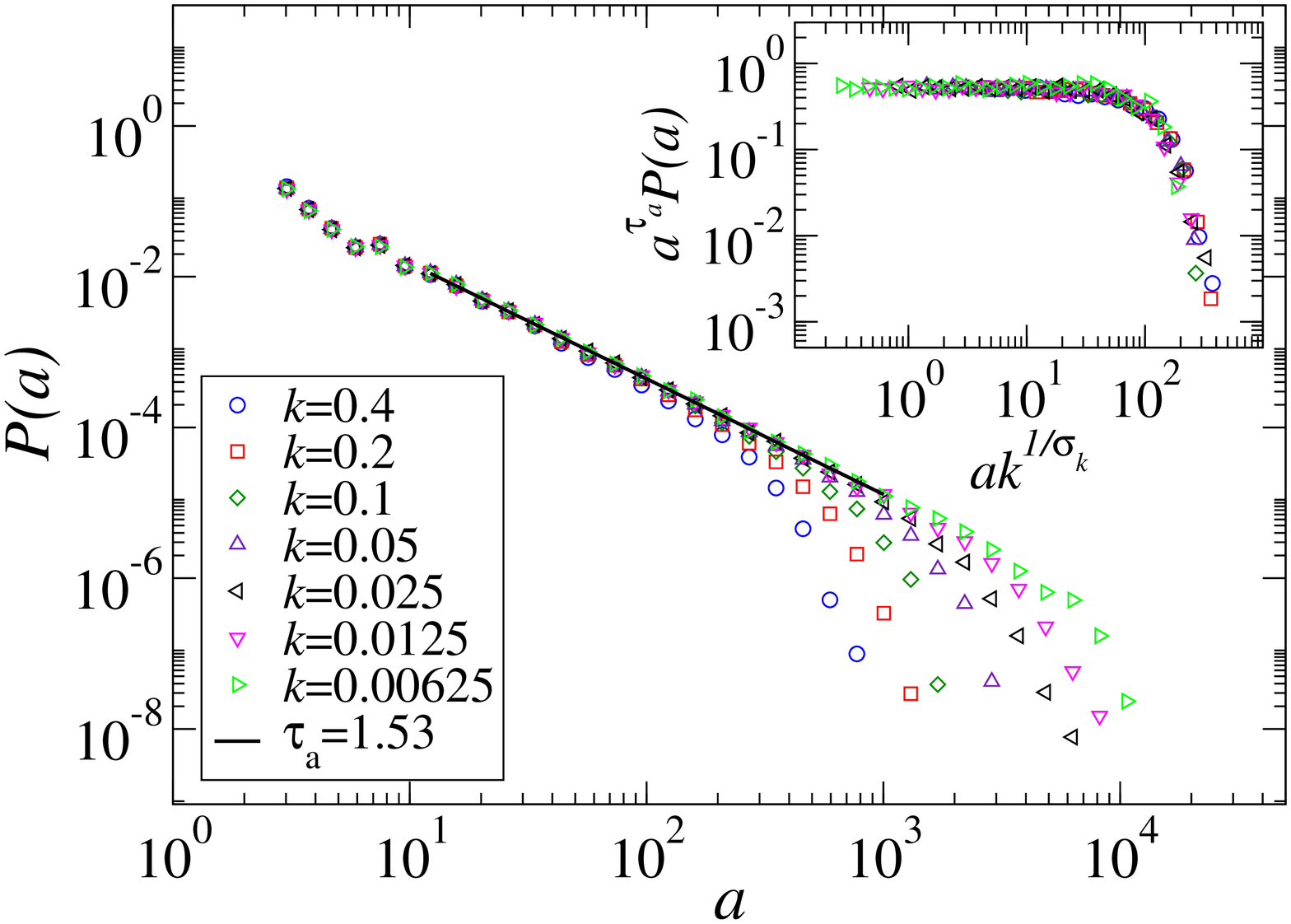}
\caption{(color online)
{\bf Top:} 
The size distributions of the AM clusters
for $\Gamma_0=1$ and various values of $k$. The solid line
is a guide to the eye and corresponds to $\tau_a=1.52$. The
inset shows a data collapse with $\tau_a=1.52$ and
$1/\sigma_k=0.7$.
{\bf Bottom:}                                                                                                            
The size distributions of the WTM clusters
for $\Gamma_0=1$ and various values of $k$. The solid line
is a guide to the eye and corresponds to $\tau_a=1.53$. The
inset shows a data collapse with $\tau_a=1.53$ and
$1/\sigma_k=0.8$.}
\label{fig:local_dists}
\end{center}
\end{figure}

\begin{figure}[ht!]
\begin{center}
\includegraphics[width=6.5cm,angle=-90,clip]
{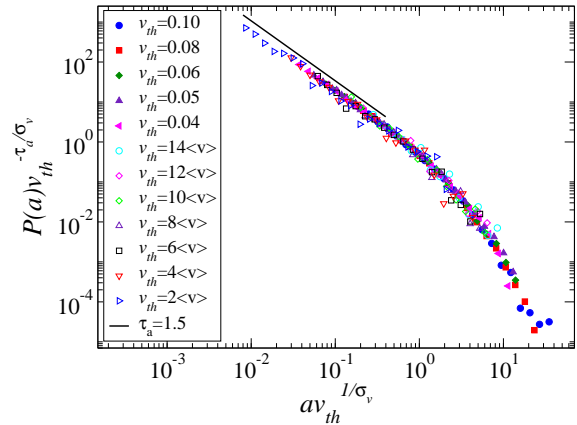}
\caption{(color online) 
The scaled size distributions of the WTM clusters
for various threshold values $v_{th}$, with $\tau_a=1.5$ and 
$1/\sigma_v=1.8$. Filled symbols correspond to the simulations, while open 
symbols are from an experiment on planar crack propagation in Plexiglas 
plates \cite{MAL-06}, with $\langle v \rangle = 11\mu m/s$ and a pixel size of 
$1.7\mu m^2$. The two data collapses (simulations and experiments) have been 
shifted on top of each other to facilitate comparison.} 
\label{fig:edists}
\end{center}
\end{figure}

\section{Results}

\subsection{Avalanche and cluster size distributions}

The probability distributions of avalanches and cluster sizes for
$L=2048$, $\Gamma_0=1$ and various values of $k$ are
shown in Figs. \ref{fig:global_dists} and \ref{fig:local_dists},
respectively. The sizes $s$ of the avalanches (Fig. 
\ref{fig:global_dists}, open symbols) are distributed according 
to a power law with a $k$-dependent cut-off, 
\begin{equation} 
P(s)=s^{-\tau}f_k(s/k^{-1/\sigma_k}),
\end{equation}
with $f_k(x)$ a scaling function describing the shape of the cut-off, 
$\tau = 1.25 \pm 0.05$ and $1/\sigma_k=0.725\pm 0.08$. 
The value of $\tau$ is in good agreement with results from 
simulations \cite{BON-08} and unpublished experiments \cite{SAN_UP}.
However, the scaling of the cut-off was not considered in Ref.~\cite{BON-08}.
Notice that the scaling function
$f_k(x)$ displays a bump around the cut-off of the distribution \cite{ROS-09}, 
and thus the true power law exponent can be seen directly only for $s$ well below 
the cut-off scale. The exponent values we quote are obtained
by optimizing the data collapse of the distributions. To demonstrate the 
robustness of our results with respect to changing the values of various 
parameters of the model, we also show the avalanche size distributions for 
a fixed $k$ but varying $\Gamma_0$ in Fig.~\ref{fig:global_dists} (filled symbols): 
the distributions turn out to be virtually independent of $\Gamma_0$ in the 
range $0.4 \leq \Gamma_0 \leq 1.0$ considered in Fig.~\ref{fig:global_dists}. 
Analogously, the cluster sizes or areas $a^{AM}$ (top panel of Fig. 
\ref{fig:local_dists}) and $a^{WTM}$ (bottom panel of Fig. 
\ref{fig:local_dists}) extracted by the AM and WTM methods, respectively,  
are both observed to scale according to
\begin{equation}
P(a)=a^{-\tau_a}
g_k(a/k^{-1/\sigma_k}),
\end{equation}
but with an exponent $\tau_a$ significantly
different from $\tau$, i.e. $\tau_a^{AM}=1.52 \pm 0.05$ and
$\tau_a^{WTM} =1.53 \pm 0.05$. Thus, the two methods to define the
clusters give the same $\tau_a$-exponent within errorbars. 
In this case the scaling function $g_k(x)$ does not exhibit a bump, and
the scaling of the distributions extends all the way
to the cut-off scale. Also for the cluster size distributions similar 
results are obtained for different values of $\Gamma_0$ (not shown).
The observed exponent value is somewhat lower than that quoted in 
Refs.~\cite{MAL-06,BON-08} for the WTM 
approach, but our analysis below indicates that also the data presented in 
\cite{MAL-06,BON-08} is perfectly consistent with $\tau_a^{WTM}\approx 1.5$. 
This is because in Refs.~\cite{MAL-06,BON-08} also the region of the
cut-offs of the distributions was included in the fits, thus overestimating
the true exponent value. 
Notice also that the cut-off of the cluster size distributions 
scales with $k$ as $k^{-1/\sigma_k}$, with 
$1/\sigma_k=0.7\pm 0.08$ for the AM approach, while we obtain
$1/\sigma_k=0.8\pm 0.08$ for the WTM. 
The different $1/\sigma_k$ exponents have the same value within
errorbars, and we suspect that the small variation observed might be
related to a finite size effect.

\subsection{Comparison with experimental results}

In order to compare directly our results to the experiments \cite{MAL-06},
we study WTM clusters as a function of a finite velocity threshold
$v_{th}$. The threshold introduces a cut-off in the distribution that 
scales as
\begin{equation}
P(a)=a^{-\tau_a}g_v(a/v_{th}^{-1/\sigma_v}),
\end{equation}
with $\tau_a=1.53\pm 0.05$ and $1/\sigma_v=1.8\pm 0.1$, see Fig. 
\ref{fig:edists}. There, we compare the simulation results with experimental
data on planar crack propagation in Plexiglas plates \cite{MAL-06}.
Both sets of data can be collapsed by using exponent values 
$\tau_a \approx 1.5$ and $1/\sigma_v \approx 1.8$.
By shifting the two data collapses on top of each other, we observe 
perfect agreement between the simulations and the experiment.
Notice again that in Refs.~\cite{MAL-06,BON-08}, the cut-offs
of the distributions were included in the power law fits, thus
overestimating the true exponent value.

\subsection{Link between global avalanches and local clusters}

\begin{figure}[t]
\begin{center}
\includegraphics[width=6.5cm,angle=-90]{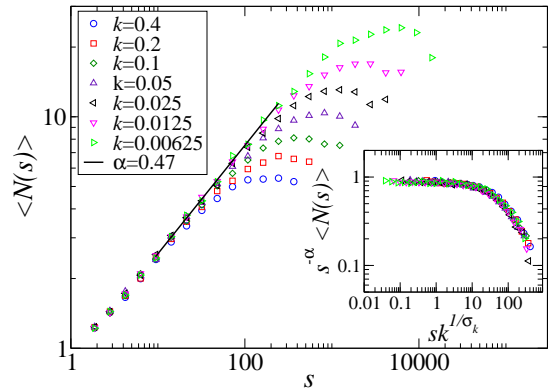}
\caption{(color online) The scaling of the average number of clusters
within a global avalanche of size $s$, for various values of $k$, and 
$\Gamma_0=1$. The solid line is a 
power law fit of the form $\langle N(s) \rangle \sim s^{\alpha}$, with 
$\alpha=0.47$. The inset shows a data collapse according to Eq. 
(\protect\ref{eq:sqrt_form}), with $\alpha=0.47$ and $1/\sigma_k=0.76$.}
\label{fig:sqrt}
\end{center}
\end{figure}

To explain these results, we consider the break-up process of an
avalanche into a number of clusters. When an avalanche grows by 
one unit (i.e. when a  single element of the line moves forward 
by one unit), there are three possibilities how this can affect 
the number $N(s)$ of distinct clusters within an avalanche: i) a 
new cluster is initiated ($N\rightarrow N+1$), ii) two existing 
clusters are merged into a single cluster ($N\rightarrow N-1$), 
or iii) an existing cluster grows by one unit ($N\rightarrow N$). 
We associate probabilities $p_1$, $p_2$ and $1-p_1-p_2$, 
respectively, to these events. These probabilities are dynamic 
variables as they depend on the instantaneous structure of the 
avalanche. On the average, however, $p_1=p_2$ as $p_1>p_2$ would 
clearly indicate an increase in $p_2$ and a subsequent tendency 
for $p_1$ to decrease again: When the number of clusters grows, 
there is less and less space to create more clusters and the 
probability to merge existing clusters should increase. Similarly,
$p_1<p_2$ would imply that sooner or later most of the clusters 
would have merged, and consequently the probability $p_2$ for 
further cluster merging decreases. Thus, for large enough 
avalanches, the number $N \geq 1$ of clusters during the growth
of an avalanche follows a random walk up to $s$ steps, with a 
reflective boundary at $N=1$ (as each avalanche consists of at 
least one cluster). Assuming for simplicity that the process is
not correlated indicates that for $s\gg 1$, 
$\langle N(s) \rangle \sim s^{\alpha}$, with $\alpha=1/2$.
Our numerical results confirm this relation, and a scaling form 
\begin{equation}
\label{eq:sqrt_form}
\langle N(s) \rangle = s^{\alpha}\tilde{N}(s/k^{-1/\sigma_k})
\end{equation}
with $\tilde{N}(x)$ a scaling function, $\alpha=0.47 \pm 0.05$ and 
$\sigma=0.76\pm 0.08$, see Fig. \ref{fig:sqrt}
(i.e. the cut-off scales with $k$ again with the same exponent $1/\sigma_k
\approx 0.75$). Thus, the cluster sizes scale as 
$a\sim s/\langle N(s)\rangle$, or $s \sim a^2$. $P(a)da=P(s)ds$ 
then implies that $P(a) \sim a^{-\tau_a}$, with
\begin{equation}
\label{eq:sr}
\tau_a = 2\tau-1.
\end{equation}
In the present case, the value of the avalanche exponent $\tau\approx 1.25$ 
yields $\tau_a \approx 1.5$ for the clusters, in good agreement
with our numerical and recent experimental results \cite{MAL-06,SAN_UP}. 
Notice that Eq.~(\ref{eq:sqrt_form}) implies that
$s \sim a^2$ holds only for $s$ smaller than the cut-off scale. For larger
avalanches one has $a \sim s$ (as evidenced by the cross-over to a plateau in
Fig. \ref{fig:sqrt}), consistent with the observation that the cut-offs of
the avalanche and cluster size distributions scale with the same exponent
$1/\sigma_k$. Notice also that while deriving Eq. 
(\ref{eq:sr}) we did not make assumptions
about the precise form of the long-range kernel nor the spatial dimension
of the system, and we thus expect Eq. (\ref{eq:sr}) to be valid for a 
wide class of avalanching systems where the avalanche dynamics is governed by 
long-range interactions. Notice that if the avalanche break-up process is 
correlated such that $\langle N(s) \rangle \sim s^{\alpha}$ with 
$\alpha \neq 1/2$, Eq. (\ref{eq:sr}) can be generalized to read 
$\tau_a=(\tau-\alpha)/(1-\alpha)$.

\begin{figure}[t!]
\begin{center}
\includegraphics[width=6.5cm,angle=-90,clip]
{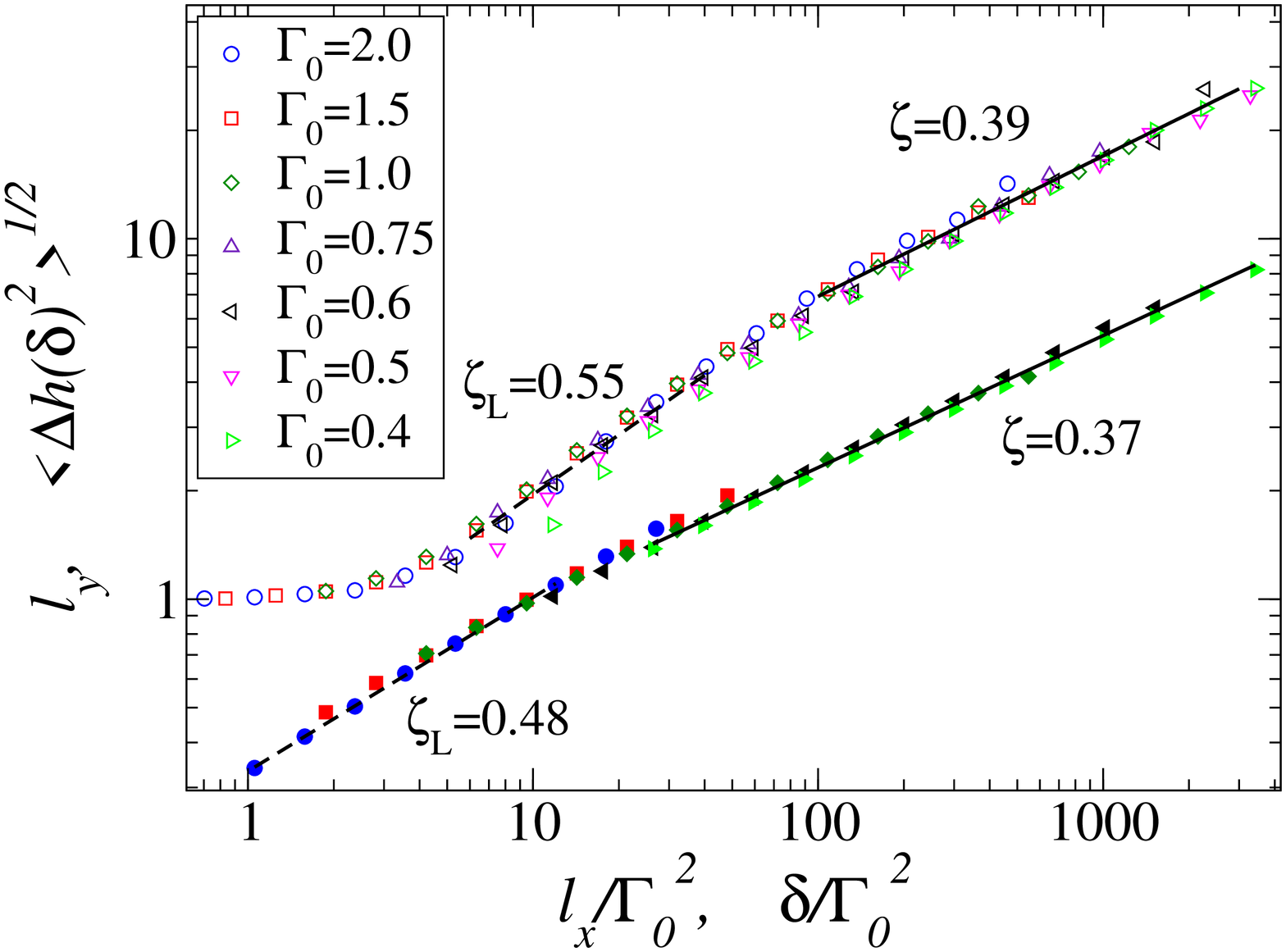}
\includegraphics[width=6.5cm,angle=-90,clip]
{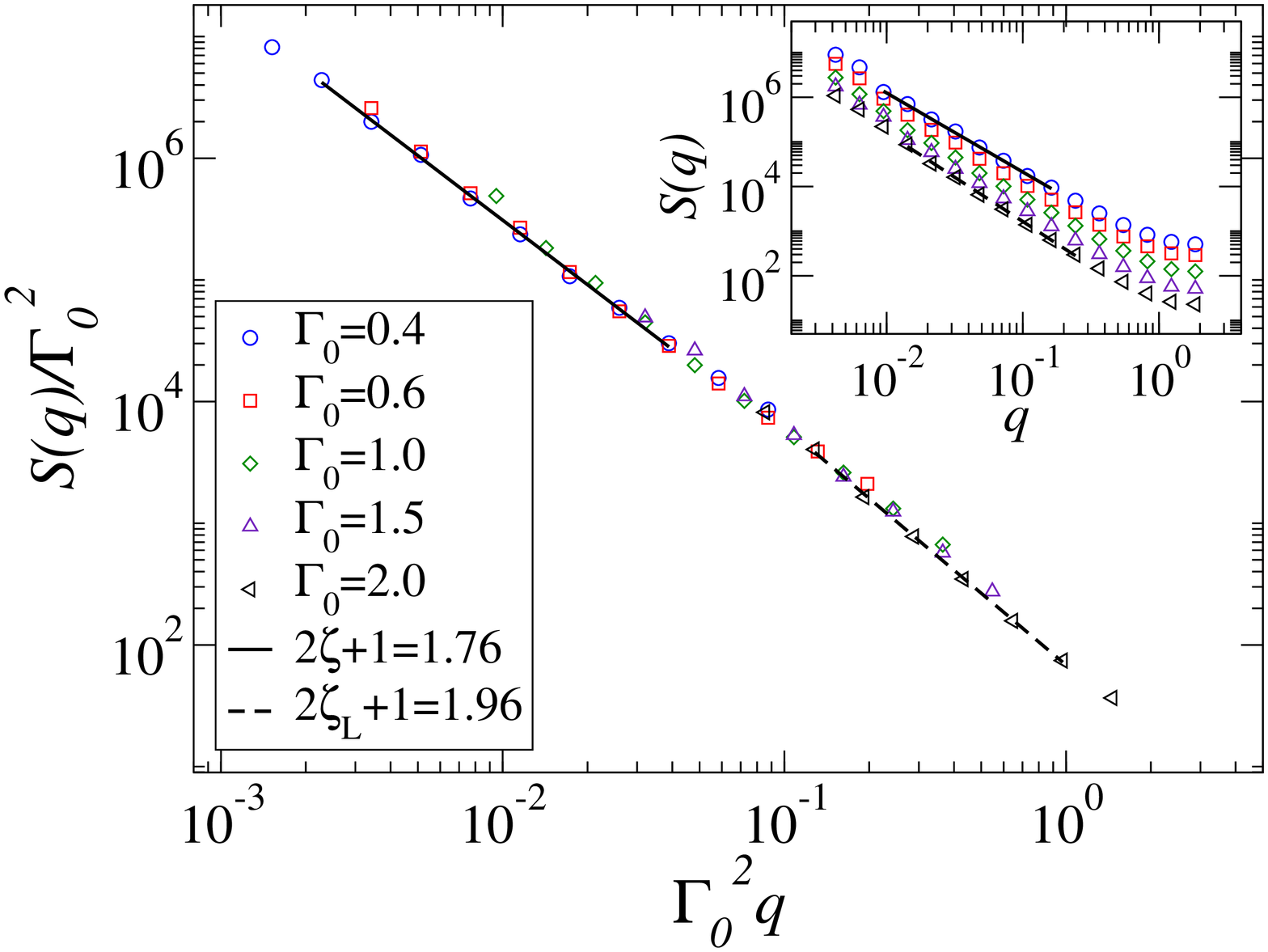}
\caption{(color online) 
{\bf Top:}
The scaling of the aspect ratio of the local clusters
(open symbols) and the root mean square height fluctuations
$\langle \Delta h (\delta)^2\rangle^{1/2}$ of the line (filled symbols),
for $k=0.0125$ and different values of $\Gamma_0$. For large clusters
their aspect ratio scales as $l_y\sim l_x^{\zeta}$, with $\zeta=0.39 \pm 0.03$ (solid line),
while smaller clusters are characterized by $\zeta_{L}\approx 0.55$
(dashed line). Also the roughness of the line, scaling like
$\langle \Delta h (\delta)^2\rangle^{1/2} \sim \delta^{\zeta}$, exhibits two distinct
scaling regimes, with $\zeta_{L} \approx 0.48$ for small scales and
$\zeta=0.37 \pm 0.03$ for large scales. Both sets of data have been collapsed by rescaling
$l_x$ and $\delta$ by the Larkin length $L_c \sim \Gamma_0^2$.
{\bf Bottom:} 
The power spectra of the line profiles for different values of $\Gamma_0$.
The main figure shows a collapse of the spectra, according to 
$S(q)=\Gamma_0^2 \tilde{S}(\Gamma_0^2 q)$, while the inset displays the 
unscaled power spectra. The power spectra scale as $S(q) \sim q^{-(2\zeta+1)}$,
with a cross-over separating regimes with $2\zeta_L+1\approx 1.96$ for short
length scales and $2\zeta+1\approx 1.76$ for long length scales.
}
\label{fig:roughness}
\end{center}
\end{figure}

\subsection{Cluster morphology and roughness of the crack front}

In experiments \cite{MAL-06} and simulations \cite{BON-08} it was found that
the aspect ratio of the clusters was scaling as $l_y \sim l_x^{\zeta}$,  with $\zeta\simeq 0.6$ and
where $l_x$ and $l_y$ are the linear sizes of the cluster in the $x$ and $y$ directions, 
respectively. This result is in agreement with the 
roughness of the crack front measured in early experiments, but disagrees with the known value of
the roughness exponent of the crack line model $\zeta=0.39$ \cite{ROS-02,DUE-07}. 
Moreover, by investigating a larger range of length scales, a recent experimental work on 
roughness of the crack front has shown a crossover from a roughness exponent of around 0.6 
to a lower value corresponding to the line model prediction of around 0.39 \cite{SAN-09,SAN-10}. 
This is surprising, since we have just shown that the experimental avalanche statistics 
is in perfect agreement with the predictions of the model. This puzzle can be resolved, 
noticing that $\zeta=0.39$ is expected to hold only for length scales larger than the Larkin 
length $L_c$ \cite{LAR-79},
which for our model scales like $L_c \sim \Gamma_0^2 \xi/R^2$, where $\xi$ and $R$ are
the correlation length and the amplitude of the disorder (here $\xi=R=1$), respectively 
\cite{MOR-04}. Below $L_c$, the crack line roughness should scale with an exponent 
$\zeta_L=1/2$. 

In the top panel of Fig.~\ref{fig:roughness} we report the
scaling of $l_y$ with $l_x$ for different values of $\Gamma_0$. By rescaling $l_x$
by $L_c\sim \Gamma_0^2$, we observe a crossover for the scaling
exponent going from $\zeta_L = 0.55 \pm 0.05$ at small scales to $\zeta=0.39\pm 0.03$ at large scales. 
In the top panel of Fig.~\ref{fig:roughness} we also report the scaling of the root mean square
height fluctuations $\langle \Delta h (\delta)^2\rangle^{1/2}$ (with $\Delta h(\delta)
=h_{i+\delta}-h_i$) of the crack front, expected to scale like 
$\langle \Delta h (\delta)^2\rangle^{1/2} \sim \delta^{\zeta}$. Again, by rescaling 
$\delta$ by $\Gamma_0^2$, we observe a crossover from a small scale regime with 
$\zeta_L =  0.48 \pm 0.05$ to the large scale value $\zeta=0.37\pm 0.03$.
The bottom panel
of Fig.~\ref{fig:roughness} shows that similar conclusion can be made by considering
the power spectrum $S(q)$ of the line profile, expected to scale as $S(q) \sim q^{2\zeta+1}$.
Rescaling the data by $L_c\sim \Gamma_0^2$ according to $S(q)=\Gamma_0^2 \tilde{S}(\Gamma_0^2 q)$,
we observe a crossover for the scaling
exponent going from $2\zeta_L+1 = 1.96 \pm 0.05$ for large $q$ to 
$2\zeta+1 \approx 1.76 \pm 0.04$ for small $q$, in agreement with the 
analysis of root mean square height fluctuations.
We have verified that in the present model, where the crack front is constrained to
move along the $y$ direction, there is no multiscaling \cite{note}.
Notice that while we observe different crossover scales for the roughness of the line
and the scaling of the morphology of the clusters, both sets of data scale 
with the Larkin length $L_c$, and thus the cross-over length is proportional to $L_c$ in 
both cases. However, the fact that these two length scales are not the same 
implies that there is a range of scales for which the roughness of the line appears
to scale with the asymptotic exponent $\zeta \approx 0.39$, while the cluster aspect
ratio is still characterized by the larger $\zeta_L \approx 0.55$. It is probably
this range of scales that has been investigated in earlier numerical studies \cite{BON-08}.

\section{Conclusions}

To summarize, we have demonstrated how avalanches are broken
into spatially disconnected clusters in systems with long range interactions. 
Both the avalanches as a whole and their localized parts or clusters
exhibit scaling, but with different power law exponents, related by a
scaling relation derived from a simple random walk argument for the 
avalanche break-up process. We also showed that large enough clusters exhibit 
aspect ratio scaling with an exponent consistent with the roughness 
exponent of the crack line model \cite{ROS-02,DUE-07}. Due to the general
nature of both the model considered as well as our arguments, we expect 
these results to be applicable in a large class of avalanching systems 
with long range interactions, ranging from crack propagation to contact 
lines in wetting to domain walls in ferromagnets. 

Of particular interest here is our explanation of the different scaling
exponents observed for global avalanches and local clusters in the 
crack line model \cite{BON-08}, and its potential implications for
further studies of avalanche phenomena in systems with long-range 
interactions. An interesting prospect would be to explore the 
possibilities to explain the observed acoustic emission exponents in 
peeling of paper sheets \cite{SAL-06} with similar arguments. The 
observation that the avalanche break-up process in the line model can 
be described by a simple random walk will be interesting to test 
in experiments on crack front propagation in Plexiglas plates, but
also in other avalanching systems with long range interactions, such
as for domain wall dynamics in ferromagnetic thin films \cite{DUR-06},
as well as for imbibition \cite{PLA-09}.

An important point regarding the roughness of the crack front and
the related cluster morphology is that their scaling properties
depend on the range of length scales considered: The asymptotic
scaling is observed only for length scales larger than a cross-over 
scale proportional to the Larkin length of the crack line. Earlier
simulations \cite{BON-08} as well as experiments \cite{MAL-06} have
most likely been probing the regime below this cross-over scale,
possibly explaining the different exponent values observed in those cases.
However, for these small length scales additional complications arise due 
to the overhangs observed in the experimental crack profiles, 
which are excluded by construction in the present line model. Such 
overhangs might be responsible for the multiscaling observed for the
experimental crack profiles for short length scales \cite{SAN-10}. 
We are working to extend the crack front line model to include the 
possibility to form overhangs, and are planning to check if such an 
extension is sufficient to reproduce the experimentally observed 
short length scale scaling properties of the crack fronts.

The connection between avalanches and clusters in avalanching systems 
with long-range interactions opens up also interesting possibilities 
to understand the observed spatio-temporal correlations between 
avalanches in a number of systems, such as the space-time coupling of 
avalanches observed in plastically deforming crystals \cite{WEI-03}, 
and even the fore- and aftershock sequences of earthquakes: One 
possibility worth further studies would be to check if such correlated 
avalanches could in fact be local clusters, with the observed correlations 
arising naturally from the fact that the localized clusters are parts 
of the same global avalanche.

{\bf Acknowledgments}. M.J. Alava and X. Illa are thanked for interesting
discussions, and  D. Bonamy for sending us the data of the avalanche 
size distributions of Ref. \cite{BON-08}. K.J. M\aa l\o y is thanked for
the possibility to publish the experimental data in Fig. \ref{fig:edists}. 
LL wishes to thank Academy of Finland for financial support.


\begin{thebibliography}{10}
\bibitem{note} 
Due to artifacts related to the discretization of the front
$y$ position $h(x)$, an effective  multiscaling is observed at very small scales, but
we have checked that this spurious effect disappears if we employ a continuous
variable for $h(x)$.
\end{thebibliography}
\end{document}